\documentstyle[12pt]{article}
\begin{document}
\centerline{\Large\bf Quantum Inequalities in Curved}
\centerline{\Large\bf Two Dimensional Spacetimes}
\vskip .8in
\centerline{Dan N. Vollick}
\centerline{Department of Mathematics and Statistics}
\centerline{and}
\centerline{Department of Physics}
\centerline{Okanagan University College}
\centerline{3333 College Way}
\centerline{Kelowna, B.C.}
\centerline{V1V 1V7}
\vskip .9in
\centerline{\bf\large Abstract}
\vskip .3in
In quantum field theory there exist states for which the energy density
is negative. It is important that these negative energy densities satisfy
constraints, such as quantum inequalities, to minimize possible
violations of causality, the second law of thermodynamics, and 
cosmic censorship. In this paper I show that conformally invariant
scalar and Dirac fields satisfy quantum inequalities in two
dimensional spacetimes with a conformal factor that depends on $x$ only
or on $t$ only. These inequalities are then applied to two dimensional
black hole and cosmological spacetimes. It is shown that the bound on
the negative energies diverges to minus infinity 
as the event horizon
or initial singularity is approached. Thus, 
neglecting back reaction, 
negative energies become 
unconstrained near the horizon or initial singularity.
The results of this paper also support the hypothesis 
that the quantum interest conjecture applies only to
deviations from the vacuum polarization energy, not to the total
energy.
\newpage
\section*{Introduction}
In classical physics it is expected that the energy-momentum tensor
will satisfy the weak energy condition $T_{\mu\nu}V^{\mu}V^{\nu}
\geq 0$ for all timelike vectors $V^{\mu}$ (see \cite{Vo1} for exceptions).
This condition ensures that all
observers measure a positive energy density. 
However, in quantum field theory there exist states for which the expectation
value of the energy-momentum tensor violates the weak energy condition.
Such exotic matter is of interest since it is required to maintain
wormholes \cite{Mo1,Mo2,Fr1,Fo4,Ho1,Vi1} and to create warp drives \cite{Al1,Pf1}.
   
Over the past decade Ford and Roman \cite{Fo1,Fo2,Fo3} have studied the 
properties of exotic matter extensively. In four dimensional flat
spacetime they have shown that massless bosonic fields satisfy the
quantum inequalities
\begin{equation}
\hat{\rho}=\frac{t_0}{\pi}\int_{-\infty}^{\infty}\frac{<T_{tt}>}
{(t^2+t_0^2)}dt\geq -\frac{3A}{32\pi^2t_0^4}
\label{qi4}
\end{equation}
where $<T_{tt}>$ is the expectation value of the normal ordered
energy density, $A=1$ for massless scalar fields, and $A=2$ for the 
electromagnetic field. The quantity $\hat{\rho}$ samples $<T_{tt}>$ over
a time interval of order $t_0$ using the sampling function
$h(t)=t_0/[\pi (t^2+t_0^2)]$. These quantum inequalities show that
an observer can measure a negative energy density for a time 
$\stackrel{<}{\sim} |\rho|^{-1/4}$. In two dimensional flat spacetime
they found that
\begin{equation}
\hat{\rho}\geq -\frac{1}{8\pi t_0^2}
\label{2d}
\end{equation}
for a massless scalar field. The bounds found by Ford and Roman
are not optimal bounds. In two dimensional flat spacetime
Flanagan \cite{Fl1} has
shown that the optimal bound for a scalar field 
is given by
\begin{equation}
\hat{\rho}\geq -\frac{1}{24\pi}\int_{-\infty}^{\infty}\frac{h^{'}
(t)^2}{h(t)}dt
\end{equation}
where $h(t)$ is any sampling function that satisfies $h(t)\geq 0$ and
$\int_{-\infty}^{\infty}h(t)dt=1$. 
Non-optimal quantum inequalities have also been found
for scalar fields in static Robertson-Walker,
de Sitter, and Schwarzschild spacetimes \cite{Pf2,Fe1,Fe2}.
  
Some work has also been done on negative energy density states for
the Dirac equation in four dimensional Minkowski space. In an earlier
paper \cite{Vo2} I examined a class of states that produced violations
of the weak energy condition and showed that they satisfied a quantum
inequality of the form (\ref{qi4}).
  
In this paper I show that (\ref{2d}) is also an optimal bound for the massless Dirac field in two dimensional flat spacetimes. I also find the optimal 
bound for massless Dirac and scalar fields in two dimensional curved
spacetimes with a conformal factor dependent on $x$ only or on $t$ only.
These results are then applied to two dimensional black hole and 
cosmological spacetimes. It is shown that the bound 
on the negative energies diverges to minus
infinity (neglecting back reaction)
as the event horizon or cosmological singularity is approached.
Thus, the negative energies become unconstrained near the horizon
or the initial singularity.
  
The quantum inequalities derived in this paper can also be used
to support the quantum interest conjecture \cite{Fo5}. According to
this conjecture any negative energy flux must be preceded or followed by
a larger positive energy flux (i.e. the negative flux must be repaid
with interest by the positive flux). Pretorius \cite{Pr1} has shown
that the quantum interest conjecture for scalar fields in Minkowski
space follows from the scaling properties of the corresponding
quantum inequalities. Since the Dirac equation satisfies the same
quantum inequalities in two dimensional Minkowski space it will also
satisfy the quantum interest conjecture. Pretorius also conjectured that
in curved spacetime the quantum interest conjecture applies only
to deviations from the ground state, not to the total energy. This conjecture
is shown to be true for the spacetimes examined in this paper.
I will take $\hbar=c=G=1$ throughout this paper.
\section*{The Energy-Momentum Tensor}
In any two dimensional spacetime the metric can be taken to have the
conformally flat form
\begin{equation}
ds^2=C(x,t)[dt^2-dx^2].
\end{equation}
In null coordinates ($u=t-x,\;\; v=t+x$) the metric becomes
\begin{equation}
ds^2=C(u,v)dudv .
\end{equation}
The conservation laws $\nabla_{\mu}T^{\mu}_{\;\;\nu}=0$ give
\begin{equation}
\frac{1}{\sqrt{g}}\frac{\partial}{\partial x^{\alpha}}\left [
\sqrt{g}T^{\alpha}_{\;\;\beta}\right ] =\frac{1}{2}T\partial_{\beta}(\ln C)
\end{equation}
where $T=T^{\alpha}_{\;\;\alpha}$ and $T^{\alpha}_{\;\;\beta}$ is the
energy-momentum tensor. Using $T^{u}_{\;\; u}=T^v_{\;\; v}=\frac{1}{2} T$
and $T_{uu}=\frac{1}{2}CT^v_{\;\; u},\; T_{vv}=\frac{1}{2}CT^u_{\;\; v}$ we
find that
\begin{equation}
\partial_v T_{uu}=-\frac{1}{4}C\partial_u T
\end{equation}
and
\begin{equation}
\partial_u T_{vv}=-\frac{1}{4}C\partial_v T .
\end{equation}
Thus in flat spacetime, where the conformal anomally vanishes (i.e. $T=0$), a 
conformally invariant theory, such as the massless scalar or Dirac field,
will have $T_{uu}=T_{uu}(u)$, $T_{vv}=T_{vv}(v)$, and $T_{uv}=0$.
This can be seen explicitly for the massless scalar and Dirac fields.
  
The scalar field equation $\Box^2\phi=0$ has the general solution $\phi
(u,v)=\phi_u(u)+\phi_v(v)$ and the energy-momentum tensor has the 
components $T_{uu}=(\partial_u\phi_u)^2,\; T_{vv}=
(\partial_v\phi_v)^2$, and $T_{uv}=0$.
  
The massless Dirac equation
\begin{equation}
i\gamma^{\mu}\partial_{\mu}\psi =0
\end{equation}
with
\begin{equation}
\gamma^0=\left (
\begin{array}{cc}
0 \;\;\;\;\; -i\\
i \;\;\;\;\;\;\; 0\\
\end{array}
\right )
\end{equation}
and
\begin{equation}
\gamma^1=\left (
\begin{array}{cc}
0 \;\;\;\;\;\;\; i\\
i \;\;\;\;\;\;\; 0\\
\end{array}
\right )
\end{equation}
has the general solution
\begin{equation}
\psi=\left [
\begin{array}{cc}
\psi_u(u)\\
\psi_v(v)\\
\end{array}
\right ] .
\end{equation}
The energy-momentum tensor is given by
\begin{equation}
T_{uu}=\frac{i}{2}[\psi_u^{\dagger}\partial_u\psi_u -(\partial_u\psi_u
^{\dagger})\psi_u],
\end{equation}
\begin{equation}
T_{vv}=\frac{i}{2}[\psi_v^{\dagger}\partial_v\psi_v-(\partial_v\psi_v
^{\dagger})\psi_v],
\end{equation}
and $T_{uv}=0$. Note that $\psi$ can be taken to be real since the gamma
matrices are purely imaginary. For a discussion of the 1+1 dimensional
Dirac equation see chapter four of Green, Schwarz, and Witten \cite{Gr1}.
Both of these energy-momentum tensors satisfy $T_{uu}
=T_{uu}(u)$, $T_{vv}=T_{vv}(v)$, and $T_{uv}=0$.
\section*{Quantum Inequalities}
Consider two conformally related spacetimes, one with metric
\begin{equation}
ds^2=C(x,t)[dt^2-dx^2]
\label{metric}
\end{equation}
and the other one with $C(x,t)=1$. The energy-momentum tensor for
scalar and Dirac fields satisfies \cite{Da1,Da3,Da2,Wa1,Bi1}
\begin{equation}
<T_{\alpha\beta}^{(C)}>=<T_{\alpha\beta}^{(\eta)}>+\Theta_{\alpha\beta}
-\frac{1}{48\pi}RC\eta_{\alpha\beta}
\label{EM}
\end{equation}
where $T_{\alpha\beta}^{(C)}$ is the renormalized energy-momentum tensor 
with the metric given in (\ref{metric}), $T_{\alpha\beta}^{(\eta)}$
is the energy-momentum tensor with $C(x,t)=1$,
\begin{equation}
\Theta_{uu}=-\frac{1}{12\pi}C^{\frac{1}{2}}\partial_u^2C^{-\frac{1}{2}} ,
\end{equation}
\begin{equation}
\Theta_{vv}=-\frac{1}{12\pi}C^{\frac{1}{2}}\partial_v^2C^{-\frac{1}{2}},
\end{equation}
\begin{equation}
\Theta_{uv}=\Theta_{vu}=0 ,
\end{equation}
and $R=\Box^2\ln C$ is the Ricci scalar.
  
First consider the quantum inequalities in a two dimensional flat spacetime.
The metric is
\begin{equation}
ds^2=dt^2-dx^2=dudv .
\end{equation}
Now consider the conformally related spacetime
\begin{equation}
ds^2=f^{'}(v)dudv = dudV
\end{equation}
where $V=f(v)$. From (\ref{EM}) we find that
\begin{equation}
<T_{vv}^{(f^{'})}>=<T_{vv}^{(\eta)}>+\Theta_{vv} ,
\end{equation}
which gives
\begin{equation}
(f^{'})^2<T_{VV}>=<T_{vv}^{(\eta)}>+\Theta_{vv} ,
\end{equation}
where $T_{VV}$ is $VV$ component of
the energy-momentum tensor in the (u,V) coordinate
system.
Now multiply by the sampling function $h(v)$ and integrate over $v$ to
get
\begin{equation}
\int_{-\infty}^{\infty}f^{'}(v)h(v)<T_{VV}>f^{'}(v)dv=\int_{-\infty}
^{\infty}h(v)<T_{vv}^{(\eta)}>dv+\int_{-\infty}^{\infty}h(v)\Theta_{vv}dv .
\end{equation}
Choose $f^{'}(v)$ such that $f^{'}(v)h(v)=1$. This gives
\begin{equation}
\int_{-\infty}^{\infty}<T_{VV}>dV=\int_{-\infty}^{\infty}h(v)<T_{vv}^{(
\eta)}>dv+\Delta
\label{int}
\end{equation}
where 
\begin{equation}
\Delta=\frac{1}{48\pi}\int_{-\infty}^{\infty}\frac{h^{'}(v)^2}{h(v)}dv.
\label{delta}
\end{equation}
Note that I have taken $h^{'}(v)\rightarrow 0$ as $v\rightarrow \pm\infty$ to 
obtain (\ref{delta}). Now the left hand side of (\ref{int}) is 
greater than or equal to zero
since it is the normal ordered Hamiltonian for the left moving 
sector. Thus
\begin{equation}
\int_{-\infty}^{\infty}h(v)<T_{vv}(v)>dv\geq -\Delta.
\end{equation}
This is an optimal bound since the equality can be reached by
using the ground state of the left moving sector. A similar result holds for 
$\int_{-\infty}^{\infty}h(u)<T_{uu}(u)>du$. Thus from $T_{tt}=
T_{uu}+T_{vv}$ we find the optimal inequality (at fixed $x$)
\begin{equation}
\hat{\rho}=
\int_{-\infty}^{\infty}h(t)<T_{tt}(t)>dt\geq -\frac{1}{24\pi}
\int_{-\infty}^{\infty}\frac{h^{'}(t)^2}{h(t)}dt
\label{rhohat}
\end{equation}
for scalar and Dirac fields in two dimensional flat spacetime. This
is the result obtained by Flanagan \cite{Fl1} for scalar fields.
As discussed in the introduction, the above quantum inequality implies 
that the Dirac field satisfies the quantum interest conjecture in
two dimensional Minkowski space.
  
Now consider a two dimensional curved spacetime with $C=C(x)$. An 
observer with velocity $V^{\alpha}$ will measure 
\begin{equation}
\rho=<T_{\alpha\beta}>V^{\alpha}V^{\beta}.
\end{equation}
for the the expectation value of the energy density.
From (\ref{EM}) we find
\begin{equation}
\rho_{(C)}=\frac{1}{C}\rho_{(\eta)}+\frac{1}{C}\left [\Theta_{\alpha
\beta}V^{\alpha}_{(\eta)}V^{\beta}_{(\eta)}-\frac{RC}{48\pi}\right ]
\label{rho}
\end{equation}
where $\rho_{(C)}$ is the energy density in the spacetime, $\rho_{
(\eta)}$ is the energy density in the conformally related flat spacetime,
and $V^{\alpha}_{(\eta)}$ is the four velocity of the observer in the
flat spacetime (note that $V^{\alpha}_{(C)}=C^{-\frac{1}{2}}V^{\alpha}_{(\eta)})$.
  
Now consider an observer at rest in the $(x,y)$ coordinate system.
Multiply
(\ref{rho}) by $h_C(t)\sqrt{C}$ and integrate to get
\begin{equation}
\int_{-\infty}^{\infty}\rho_{(C)}h_C\sqrt{C}dt=\frac{1}{C}\int
_{-\infty}^{\infty}\rho_{(\eta)}h_{\eta}dt-\hat{\Delta}
\label{eqn}
\end{equation}
where $h_{\eta}=\sqrt{C}h_C$ and
\begin{equation}
\hat{\Delta}=-\frac{1}{C}\left [\Theta_{tt}-\frac{RC}{48\pi}\right ].
\end{equation}
Note that $\int_{-\infty}^{\infty}h_C\sqrt{C}dt=\int_{-\infty}
^{\infty}h_{\eta}dt=1$. Using $R=\Box^2\ln C$ and $\Theta_{tt}=
-\frac{1}{24\pi}C^{\frac{1}{2}}\partial_x^2(C^{-\frac{1}{2}})$ gives
\begin{equation}
\hat{\Delta}=\frac{1}{6\pi C}[C^{\frac{1}{4}}\partial^2_x(C^{-\frac{1}{4}})].
\label{deltahat}
\end{equation}
Thus from (\ref{rhohat}), (\ref{eqn}), and (\ref{deltahat}) we find that
\begin{equation}
\hat{\rho}_C\geq-\frac{1}{24\pi C}\left [\int_{-\infty}^{\infty}
\frac{h^{'}_{\eta}(t)^2}{h_{\eta}(t)}dt+4C^{\frac{1}{4}}\partial^2_x(C^{-
\frac{1}{4}})\right ].
\label{rho2}
\end{equation}
This is an optimal bound since the equality can be reached by using the
state that minimizes $\int_{-\infty}^{\infty}\rho_{(\eta)}h_{\eta}dt$.
The first term on the right hand side of (\ref{rho2}) will satisfy the
scaling arguments used by Pretorius \cite{Pr1}, but the second term (i.e. the 
vacuum polarization term) does not. Thus, the quantum interest conjecture
applies to deviations from the vacuum polarization energy, not to the
total energy.
  
A similar bound can be derived for spacetimes with $C=C(t)$. In this case
we will average over space instead of time. It is easy to show that 
(\ref{rhohat}) becomes
\begin{equation}
\tilde{\rho}=\int_{-\infty}^{\infty}h(x)<T_{tt}>dx\geq -\frac{1}{24\pi}
\int_{-\infty}^{\infty}\frac{h^{'}(x)^2}{h(x)}dx
\end{equation}
and that (\ref{rho2}) becomes
\begin{equation}
\tilde{\rho}_C\geq -\frac{1}{24\pi C}\left [\int_{-\infty}^{\infty}
\frac{h^{'}_{\eta}(x)^2}{h_{\eta}(x)}dx+\frac{1}{2}\left (\frac{
\partial_t C}{C}\right )^2\right ].
\label{cos}
\end{equation}
This is also an optimal bound since the equality can be reached by using
the state that minimizes $\int_{-\infty}^{\infty}\rho_{\eta}h_{\eta}dx$.
From the above inequalities we see that the right hand side will
generally diverge as $C\rightarrow 0$. Thus, neglecting back reaction,
the negative energy densities become unconstrained in regions where
$C\rightarrow 0$.
\section*{Black Hole and Cosmological Spacetimes}
Consider the two dimensional black hole spacetime
\begin{equation}
ds^2=\left (1-\frac{2m}{\bar{r}}\right ) dt^2-\left ( 1-\frac{2m}{\bar{r}}
\right )^{-1}d\bar{r}^2 .
\end{equation}
In conformally flat coordinates
\begin{equation}
ds^2=\left [1-\frac{2m}{\bar{r}(r)}\right ](dt^2-dr^2)
\end{equation}
where $\bar{r}(r)$ is defined via
\begin{equation}
r=\bar{r}+2m\ln \left |\frac{\bar{r}}{2m}-1\right |.
\end{equation}
Note that the horizon is at $\bar{r}=2m$ and at $r\rightarrow -\infty$.
  
Now consider an observer at rest near the horizon. Close to the
horizon
\begin{equation}
C(r)\simeq e^{r/2m}
\end{equation}
and
\begin{equation}
\hat{\rho}_C\geq -\frac{1}{24\pi C}\left [\int_{-\infty}^{\infty}
\frac{h^{'}_{\eta}(t)^2}{h_{\eta}(t)}dt+\frac{1}{16m^2}\right ].
\label{ineq1}
\end{equation}
Thus, neglecting back reaction, as  
the horizon is approached the right hand side
diverges and the quantum inequality does not constrain the negative 
energy densities.
  
Consider a specific example using the sampling function
\begin{equation}
h_C(t)=\frac{\tau_0}{\pi[Ct^2+\tau_0^2]},
\end{equation}
which has been used by Ford and Roman in the flat spacetime
case with $C=1$. Note that $\sqrt{C}t$ is the proper time measured
by the observer. The function has a maximum height of $1/(\pi\tau_0)$
and a proper width of $\sim\tau_0$. The flat spacetime sampling
function is
\begin{equation}
h_{\eta}=\frac{\sqrt{C}\tau_0}{\pi[Ct^2+\tau_0^2]}.
\end{equation}
It has a maximum height of $\sqrt{C}/(\pi\tau_0) (\rightarrow 0$ as 
$\bar{r}\rightarrow 2m$) and a proper width of $\sim \tau_0/\sqrt{C}
(\rightarrow\infty$ as $\bar{r}\rightarrow 2m$). The inequality
(\ref{ineq1}) becomes
\begin{equation}
\hat{\rho}_C\geq -\frac{1}{48\pi C}\left [\frac{C}{
\tau_0^2}+\frac{1}{8m^2}\right ].
\end{equation}
  
Let's see how the Boulware state near the horizon compares with the above
inequality. Since the Boulware state diverges at the event horizon it
will not be the quantum state outside a black hole. It is instead the
quantum state exterior to a static mass distribution that is larger than
its Schwarzschild radius. I will take the mass distribution to be only
slightly larger than its Schwarzschild radius (of course, the matter
will be under arbitrarily large stresses which is physically unrealistic).
For a metric of the form
\begin{equation}
ds^2=f(\bar{r})dt^2-\frac{1}{f(\bar{r})}d\bar{r}^2
\end{equation}
the energy density in the Boulware state is given by \cite{An1}
\begin{equation}
\rho_B=\frac{1}{24\pi}\left [ f^{''}-\frac{(f^{'})^2}{4f}\right ].
\end{equation}
Note that this will be the energy density in the $(r,t)$ coordinate
system since $\rho=-T^t_{\;\; t}$ is unaffected by changes in the
radial coordinate.
For $f=(1-2m/\bar{r})$ we have
\begin{equation}
\rho_B=-\frac{1}{384\pi m^2}\left ( 1-\frac{2m}{\bar{r}}\right )^{-1}
\end{equation}
near the horizon. This is just the second term on the left
hand side of (\ref{ineq1}). Thus the Boulware state satisfies the
quantum inequality (\ref{ineq1}), as it must.
  
Now consider the cosmological spacetime
\begin{equation}
ds^2=t^n\left [dt^2-dx^2\right ]\;\;\;\;\;\;\;\;\; n>0 .
\end{equation}
The last term on the right hand side of (\ref{cos}) is
\begin{equation}
-\frac{1}{48\pi C}\left (\frac{\partial_t C}{C}\right )^2=-\frac{n^2}
{48\pi}t^{-(n+2)}
\end{equation}
which diverges as $t\rightarrow 0$. Thus as we approach the initial 
singularity the negative energy densities become unconstrained.
Of course, as one approaches the initial singularity quantum gravity
effects are expected to become important and the above analysis will
break down.
\section*{Conclusion}
Optimal quantum inequalities were derived for scalar and Dirac fields in 
two dimensional
spacetimes in which the conformal factor is a function of $x$ only or of
$t$ only. 
For spacetimes with a metric of the form
\begin{equation}
ds^2=C(x)(dt^2-dx^2)
\end{equation}
it was shown that the energy density satisfies the optimal bound
\begin{equation}
\hat{\rho}=\int_{-\infty}^{\infty}h(t)<T_{tt}>dt\geq -\frac{1}{24\pi C}
\left [\int_{-\infty}^{\infty}\frac{h^{'}(t)^2}{h(t)}dt+4C^{\frac{1}{4}}
\partial_2^x(C^{-\frac{1}{4}})\right ]
\end{equation}
where h(t) is a sampling function that satisfies $h(t)\geq 0$ and
$\int_{-\infty}^{\infty}h(t)dt=1$. A similar quantum inequality
was also found when the conformal factor is a function of time only.
These inequalities were then applied to black hole and cosmological 
spacetimes.
It was shown that the bound on the negative
energies diverges to minus infinity as the horizon or initial
singularity is approached. Thus, neglecting back reaction,
the negative energy densities become
unconstrained near the horizon or initial singularity.
  
I also showed that the quantum interest conjecture holds for scalar
and Dirac fields in two dimensional
Minkowski space and that in curved spacetimes
this conjecture applies
to deviations from the vacuum polarization energy, not to the total energy.
\section*{Acknowledgements}
I would like to thank Werner Israel, Shinji Mukohyama, Frans
Pretorius, and Thomas Roman
for many interesting discussions on negative energy densities.

\end{document}